\documentstyle[epsf,12pt]{article}
\begin{document}
\rightline{TMUP-HEL-9404}
\rightline{March 1994}
\def\pp{\vec{\relax{\kern .15em  p}}}
\baselineskip=18pt
\vskip 0.5in
\begin{center}
{\large{\bf A TEST OF A KIND OF THE EQUIVALENCE PRINCIPLE}}
{\large{\bf BY LONG-BASELINE NEUTRINO-OSCILLATION EXPERIMENTS\footnote
{Talk given at the Workshop on General Relativity and Gravitation held
at University of Tokyo, January 17-20, 1994.}}}
\end{center}
\vskip 0.2in
\begin{center}
Osamu Yasuda\footnote{Work partially supported by Grant-in-Aid for
Scientific Research of the Ministry of Education, Science and Culture
\# 05740183 and \# 05302016}
\vskip 0.2in
{\it Department of Physics, Tokyo Metropolitan University}

{\it 1-1 Minami-Osawa Hachioji, Tokyo 192-03, Japan}
\end{center}

\vglue .5in

\baselineskip=14pt
In this talk I discuss neutrino oscillations due to
possible nonuniversality of the gravitational coupling constants.
It has been pointed out \cite{Gasperini} \cite{Halprin} that a breakdown of the
universality of the gravitational couplings to different neutrino flavors
could lead to neutrino oscillations.  In particular the authors of
\cite{Halprin} studied the possibility in which the solar neutrinos
can be used to test this kind of breakdown of the
universality.  Since the flux of the solar neutrinos is relatively small
and the energy spectrum is beyond our control, the utility of the solar
neutrino for this purpose is limited.  Here we would like to propose a possible
long-baseline experiments of neutrino oscillations to test the breakdown of
the universality of the gravitational couplings to neutrinos \cite{IMY}
\cite{IMY2} \cite{PHL}.
As we will see, the breakdown of the universality of
the gravitational couplings to neutrinos of different flavors leads to a
violation of Einstein's equivalence principle (see, e.g., \cite{Gravitation})
which states that all the laws of physics must take on their familiar
special-relativistic forms in any and every local Lorentz frame, anywhere and
any time in the universe.  No consistent theory is known to predict
such nonuniversality of the gravitational coupling constants, so our motivation
is to give an upper bound on such nonuniversality.
In the present case, it turns out that we can probe the
magnitude of the breakdown of Einstein's equivalence principle to the order
of $10^{-15}$, assuming that there are neutrino mixings.  Among various
experiments to test the equivalence principle (see e.g., Ref. \cite{Will}
for a review), there have been few tests of Einstein's equivalence principle
for neutrinos \cite{Longo}.  The universality of the gravitational couplings
that I discuss in this talk is of different type from these experiments in
the past, so our discussions here are complementary to them.

In this talk we assume that there are two neutrino flavors which have
different couplings to gravity and that the eigenstates of these
different gravitational couplings do not coincide with those of the
electroweak flavors.  Throughout the present discussions we consider
neutrino oscillations between two flavors for simplicity.  Let us
suppose that two kinds of neutrinos are in a certain background field
with the different gravitational couplings, and that they are described by
the following Lagrangian
\newpage
\begin{eqnarray}
{\cal L}=&~&e(G_1){\overline \nu_1}\left[ ie^{a\mu}(G_1)
\gamma_a D_\mu(G_1)-m_1\right]\nu_1\nonumber\\
&+&e(G_2){\overline \nu_2}\left[ ie^{a\mu}(G_2)
\gamma_a D_\mu(G_2)-m_2\right]\nu_2\nonumber\\
&+& {\rm (interaction~ terms~ with~ electroweak~ gauge~ bosons)},
\label{eqn:Lagrangian}
\end{eqnarray}
where we have included mass terms to keep generality,
$e^{a\mu}(G_i)~(i=1,2)$ are the vierbein fields of some background metric
with different Newton constants $G_i~(i=1,2)$, and $e(G_i)\equiv\det
e^a_\mu(G_i)$.  For simplicity we assume that the eigenstates of the
gravitational couplings coincide with those of the masses.  Notice that
even if these neutrinos are massless, we cannot rotate these two fields
so that these are the eigenstates of the electroweak theory, since
the gravitational coupling terms are not invariant under the rotation
in the flavor space.  Since the gravitational couplings for these two kinds of
neutrinos are different, even if we choose a coordinate system in which
the Dirac equation for $\nu_1$ in (\ref{eqn:Lagrangian}) becomes the one
in a flat space-time, the Dirac equation for $\nu_2$ in the same coordinate
system does not necessarily do so.  Thus Einstein's equivalence principle
is violated in (\ref{eqn:Lagrangian}).

The configuration of the long-baseline experiment we will discuss is
depicted in Fig. 1, and the neutrino beams go underneath the ground.
The density of the Earth is not constant, so in principle we have to
regard the density as a function of the radius \cite{Stacey}.  When we
consider the case like the longbase-line experiment in the DUMAND
project \cite{Injector}, where the neutrino beam goes deep in the
Earth ($L\equiv$(length of the trajectory of the neutrino beam)$\sim$
6,000 km), this is actually the
case.  However, in the case like the one in the SOUDAN2 project
\cite{Injector}, where the neutrino beam goes just slightly under the
surface of the Earth ($L\sim$ 800 km), we can assume that the density
is approximately constant.  In this talk we will take this assumption
for simplicity, and the analysis without this approximation is given
in \cite{IMY2}.
\begin{center}
\hglue -1truecm \epsfysize=160pt \epsfbox{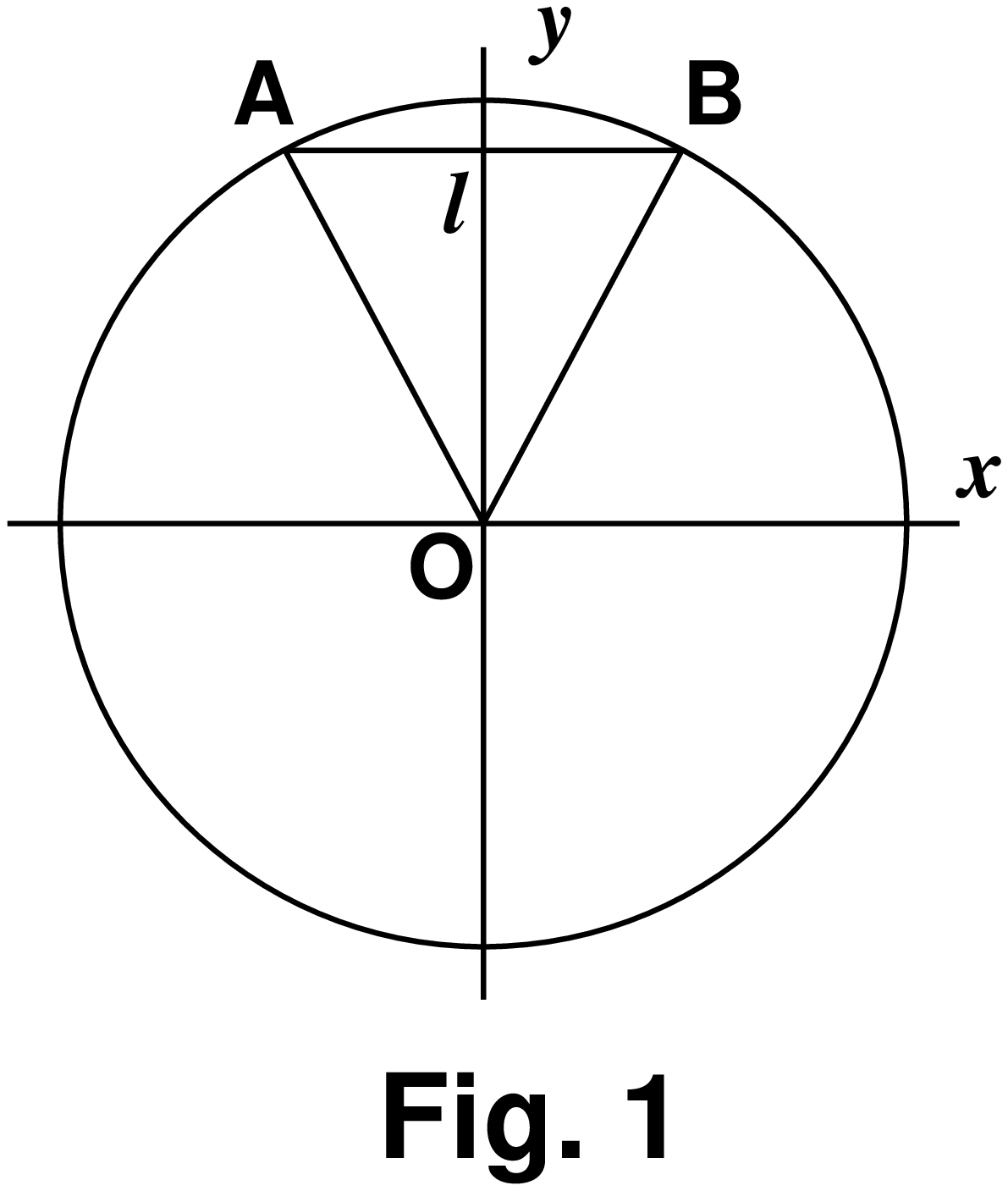}
\end{center}

First let us
consider the Dirac equation of left-handed neutrinos without any flavor
in the interior Schwarzschild background \cite{Moller}:

\begin{eqnarray}
 (ie^{a\mu}\gamma_a D_{\mu} - m)\psi = 0,
\label{eqn:Dirac1}
\end{eqnarray}
where $e_{a\mu}$ is the vierbein of the interior Schwarzschild metric
\begin{eqnarray}
ds^2 = (e^0_t)^2 dt^2 - (e^1_r)^2 dr^2 - (e^2_\theta)^2 d\theta^2
- (e^3_\phi)^2 d\phi^2
\label{eqn:metric}
\end{eqnarray}
and is given by
\begin{eqnarray}
e^0_t = {3 \over 2}\sqrt{1-{\alpha \over R}} - {1 \over 2}
\sqrt{1-{\alpha r^2 \over R^3}},~~e^1_r = {1 \over
\sqrt{1-{\alpha r^2 \over R^3}}},
{}~~e^2_\theta = r,~~e^3_\phi = r\sin\theta.
\label{eqn:vierbein}
\end{eqnarray}
$D_\mu\psi\equiv(\partial_\mu -{1 \over 2}\omega_{\mu ab}\sigma^{ab})\psi$
is the covariant derivative acting on a spinor $\psi$, $\omega_{\mu ab}$
is the spin connection given by $e^b_{[\nu}\omega^a_{\mu ]b} = \partial_{[\mu}
e^a_{\nu]}$, and $\alpha$ in (\ref{eqn:vierbein}) is the Schwarzschild
radius.

One characteristic dimensionless parameter in our case is $ER$, where $E$ is
the energy of the neutrino, and $R$ is the radius of the Earth.  For
$E$=10 GeV and $R$=6,400 Km, $ER\sim 3\times10^{23}$, and derivative
terms in the spin connections are all of the order of $1/ER$, so we
will neglect them throughout this talk.  $\alpha$ in eq.
(\ref{eqn:vierbein}) is the Schwarzschild radius of the Earth which is
about 9 mm, so we also expand (\ref{eqn:Dirac1}) to the first order in
$\alpha/r$.  In this approximation the positive energy part of the
Dirac equation finally becomes
\begin{eqnarray}
{d\nu \over du}=
i\left[ 1-{m^2 \over 2E^2}+{3\alpha \over 4R}-{\alpha \ell^2 \over 4R^3}
\left( 1-{u^2 \over B^2}\right) \right] \nu,\nonumber
\end{eqnarray}
where $u\equiv Ex$ is a dimensionless coordinate along the $x$-axis,
$\ell\equiv
\sqrt{R^2-(L/2)^2}$ is the distance of the trajectory of the neutrino
beam from the center of the Earth, and $B\equiv E\ell$ is a very large number.

In case of two kinds of neutrinos described by the Lagrangian
(\ref{eqn:Lagrangian}), it is
straightforward to see that the Dirac equation for
(\ref{eqn:Lagrangian}) is given by
\begin{eqnarray}
{d{~} \over du}\left(
\begin{array}{cc}
\nu_1 \\ \nu_2
\end{array}
\right)
=i\left[1-{m^2_1+m^2_2 \over 4E^2}+
\left( -3+{\ell^2 \over R^2}
\left( 1-{u^2 \over B^2}\right)\right){f_1+f_2 \over 4}\Phi
+\Delta(u)
\sigma_3\right] \left(
\begin{array}{cc}
\nu_1 \\ \nu_2
\end{array}
\right),
\label{eqn:Dirac2}
\end{eqnarray}
where
\begin{eqnarray}
\Delta(u)\equiv {\Delta m^2 \over 4E^2}+\left(
3-{\ell^2 \over R^2}
\left( 1-{u^2 \over B^2}\right){\Delta f \over 4}\Phi\right) .
\end{eqnarray}
Here $\Delta m^2\equiv m^2_2-m^2_1$ is the
difference of the masses, we have defined the Newton potential
$\Phi\equiv -GM/R$ on the surface of the Earth, and we have also defined
the difference $\Delta f=f_2-f_1$ of the dimensionless gravitational
couplings of the two neutrino species
\begin{eqnarray}
\left(
\begin{array}{cc}
f_1\Phi \\ f_2\Phi
\end{array}
\right)
=-\left(
\begin{array}{cc}
{\alpha_1/2R} \\ {\alpha_2/2R}
\end{array}
\right)
=-\left(
\begin{array}{cc}
{G_1M/R} \\ {G_2M/R}
\end{array}
\right) .\nonumber
\end{eqnarray}
The equation (\ref{eqn:Dirac2}) can be easily integrated from $u=-EL/2$
to $u=EL/2$.

Now let us introduce the flavor eigenstates $\nu_a,~\nu_b$ of the weak
interaction by
\begin{eqnarray}
\left(
\begin{array}{cc}
\nu_a \\ \nu_b
\end{array}
\right)
=
\left(
\begin{array}{cc}
\cos\theta&-\sin\theta\\
\sin\theta&\cos\theta
\end{array}
\right)
\left(
\begin{array}{cc}
\nu_1\\ \nu_2
\end{array}
\right) .\nonumber
\end{eqnarray}
Then the probability of detecting a different flavor $\nu_b$ at a distance
$L$ after producing one neutrino flavor $\nu_a$ is given by
\begin{eqnarray}
P(\nu_a\rightarrow\nu_b)=\sin^22\theta\sin^2\left[
\left( {\Delta m^2 \over 4E^2} + \left(
1+{L^2 \over 6R^2}\right){\Delta f\Phi \over 2}
\right) EL\right] .
\label{eqn:prob1}
\end{eqnarray}
This formula applies to the transition between $\nu_\mu$ and $\nu_\tau$,
where no MSW effect \cite{MSW} is expected to occur.

In case of the transition between $\nu_e$ and $\nu_\mu$, we have to take
the MSW effect \cite{MSW} into consideration, and the Dirac equation
is modified as
\begin{eqnarray}
{d{~} \over du}\left(
\begin{array}{cc}
\nu_e\\ \nu_\mu
\end{array}
\right) =i\left(
\begin{array}{cc}
\Delta(u)\cos 2\theta-{G_FN_e \over \sqrt{2}E}&\Delta(u)\sin 2\theta\\
\Delta(u)\sin 2\theta&-\Delta(u)\cos 2\theta+{G_FN_e \over \sqrt{2}E}
\end{array}
\right) \left(
\begin{array}{cc}
\nu_e\\ \nu_\mu
\end{array}
\right) ,
\label{eqn:Dirac3}
\end{eqnarray}
where $G_F$ is the Fermi coupling constant, $N_e$ is the density of electrons
in the Earth.
(\ref{eqn:Dirac3}) can be solved in the same way as before by introducing
the variables
\begin{eqnarray}
\Delta_N(u)\cos 2\theta_N&=&\Delta(u)\cos 2\theta-{G_FN_e \over \sqrt{2}E}
\nonumber\\
\Delta_N(u)\sin 2\theta_N&=&\Delta(u)\sin 2\theta.
\label{eqn:DeltaN}
\end{eqnarray}
Note that $\theta_N$ does depend on the variable $u$ in this case.
It is easy to integrate (\ref{eqn:DeltaN}), and we have the transition
probability of detecting $\nu_e$ at a distance $L$ from the source of
$\nu_\mu$ beams
\begin{eqnarray}
P(\nu_\mu\rightarrow\nu_e)=\sin^2 2\theta_N\left( u={EL \over 2}\right)
\sin^2\left( \int^{EL/2}_{-EL/2}du\,\Delta_N(u)\right),
\label{eqn:prob2}
\end{eqnarray}
where we have used the fact $\theta_N(u={EL \over 2})=\theta_N(u=-{EL
\over 2})$, and $\Delta_N$, $\theta_N$ are defined through (\ref{eqn:DeltaN}).
The integration in the exponent in (\ref{eqn:prob2}) can be performed
numerically \cite{IMY}.

We have studied the quantity $P(\nu_a\rightarrow\nu_b)$ for various
cases \cite{IMY} \cite{IMY2}.  As in the case of neutrino oscillations
due to masses, we can exclude certain regions in the ($\sin^2 2\theta$,
$\Delta f$) plot.  A typical figure is given in Fig. 2, which shows
the region ($\sin^2 2\theta$, $\Delta f$) for $\nu_\mu$-$\nu_\tau$
oscillations which can be excluded by the proposed
longbase-line experiment in the SOUDAN2 project.  From Fig. 2 we see
that we can give an upper limit as small as 10$^{-14}$ on the
nonuniversality of the gravitational coupling constants of neutrinos.
In case of the DUMAND project this upper limit could reach 10$^{-15}$
\cite{IMY2}.
\vglue -0.3truecm
\begin{center}
\hglue -1truecm \epsfysize=228pt \epsfbox{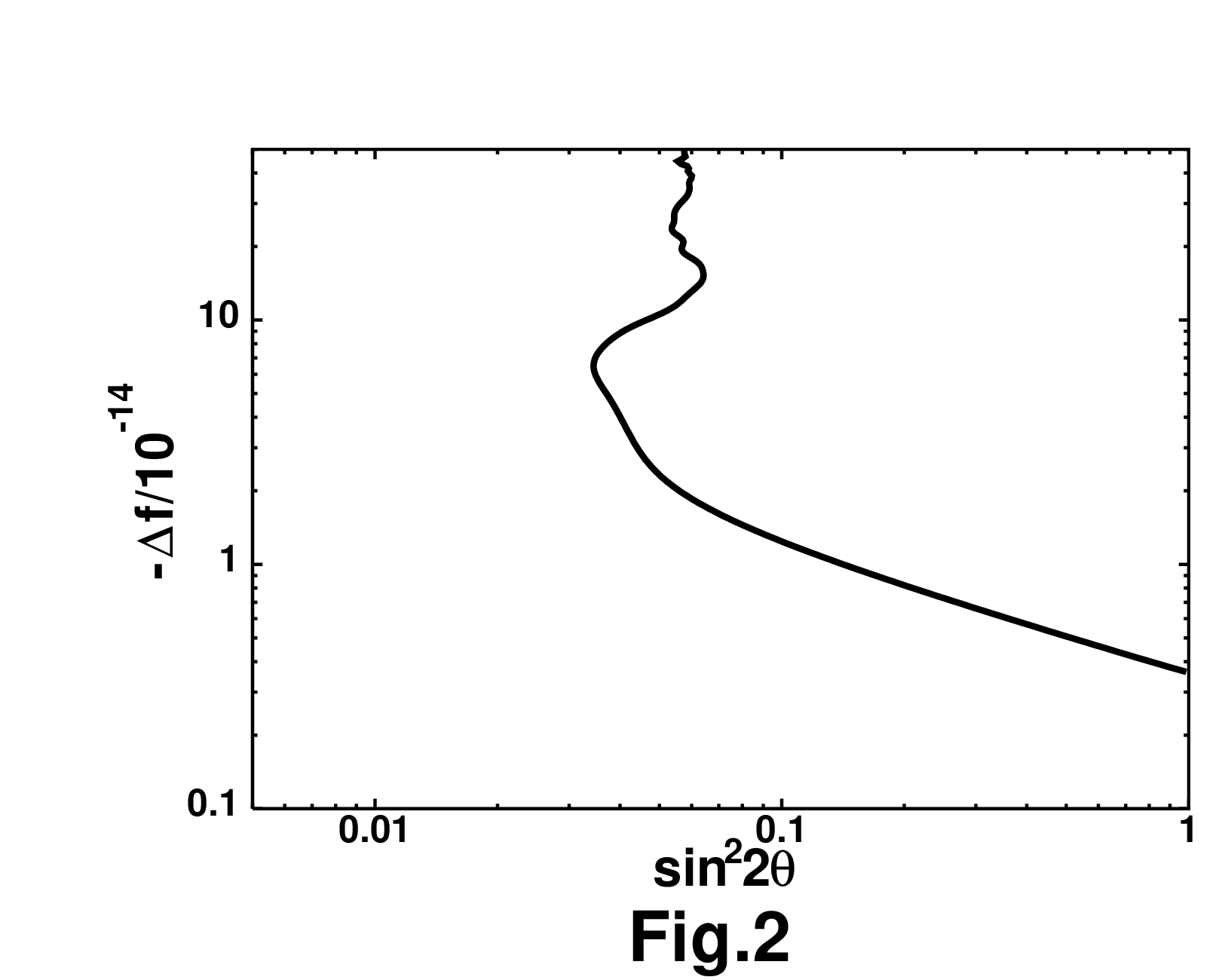}
\end{center}
\newpage
One important feature of this experiment is the dependence of
the probability on the detection threshold energy of muons,
which is, roughly speaking, proportional to the energy of the
incident neutrinos.  To emphasize this aspect, let us make
comparison of the energy dependence of the probability for
various cases.
So far we have considered the case where the particle which
mediates the force between neutrinos is a spin-2 particle, i.e.,
graviton.  In this case the energy defined in the Dirac equation
\begin{eqnarray}
i{\partial\psi \over \partial t} = E\psi
\end{eqnarray}
is given by
\begin{eqnarray}
E_{J=2}=\left( 1+{1 \over 2}\Phi_{J=2}\right)\sqrt{\pp^2+m^2}
\label{eqn:J-2}
\end{eqnarray}
On the other hand, if the force is mediated by a scalar or a vector
particle, then the energy would be given by
\begin{eqnarray}
E_{J=0}&=&\sqrt{\pp^2+(m+\Phi_{J=0})^2}\nonumber\\
E_{J=1}&=&\sqrt{\pp^2+m^2}+\Phi_{J=1},
\label{eqn:J-01}
\end{eqnarray}
respectively.  Here $\Phi_J~(J=0,1,2)$ denotes certain potentials for
spin-J forces.  Examples for such forces are a scalar field in
\cite{KMY} which couples only to tau neutrinos and a torsion tensor
field (i.e., a dual of an axial vector filed) in \cite{torsion} whose
eigenstates are different from those of the electroweak interaction.
It is well-known that the nonrelativistic behavior of scalar forces is
similar to that of tensor forces, i.e., gravity.  In the
ultrarelativistic limit, however, the situation changes drastically.
In this limit we have from (\ref{eqn:J-2}) and (\ref{eqn:J-01})
\begin{eqnarray}
\nonumber\\
E_{J=0} - \vert \pp \vert-{m^2 \over 2\vert\pp\vert} &\simeq &
{1 \over 2\vert \pp\vert}\Phi_{J=0}^2+{m \over \vert \pp\vert}\Phi_{J=0}
\nonumber\\
E_{J=1} - \vert \pp \vert-{m^2 \over 2\vert\pp\vert} &\simeq &
\Phi_{J=1}
\nonumber\\
E_{J=2} - \vert \pp \vert-{m^2 \over 2\vert\pp\vert} &\simeq &
{\vert \pp\vert \over 2}\Phi_{J=2},
\end{eqnarray}
where only terms proportional to $m$ and $\Phi_J$ could be relevant to
neutrino oscillations.  So if all neutrinos are messless and if
neutrino oscillations occur solely due to the presence of $\Phi_J$,
then the probability of neutrino oscillations would be given by
\begin{eqnarray}
{\rm Prob}(J=0) &\sim &\sin^2 2\theta \sin^2\left(
{\Phi_{J=0}^2L \over  \vert \pp \vert}\right)
\nonumber\\
{\rm Prob}(J=1) &\sim &\sin^2 2\theta \sin^2\left(
\Phi_{J=1}L\right)
\nonumber\\
{\rm Prob}(J=2) &\sim &\sin^2 2\theta \sin^2\left(
\vert \pp\vert\Phi_{J=2}L\right).
\label{eqn:Prob012}
\end{eqnarray}
\newpage
In Fig. 3 we give the energy dependence of the probability in case of
gravity.  In case of a scalar force and a vector force the energy
dependence would look like Figs. 4 and 5, respectively.  From these
figures we conclude that the shape of the energy spectrum depends on
the spin of the particle which mediates the force, and hence it should
be possible to identify the spin of the particle which mediates a
possible new force by looking at the energy dependence.
(\ref{eqn:Prob012}) also explains why we have obtained such a severe
constraint on the universality of the gravitational coupling constants
of neutrinos, as the argument of sine in Prob($J$=2) in
(\ref{eqn:Prob012}) is proportional to the energy of the neutrino.  We
note in passing that ordinary neutrino oscillations due to masses is
analogous to the case of a scalar force, as far as the energy
dependence is concerned.  In accelerator experiments like those we
have proposed in this talk, the length $L$ is much smaller and the
energy of neutrinos $\vert\pp\vert$ is much larger than typical
quantities in astrophysical observations discussed in
\cite{Will}.  In longbase-line experiments, therefore, while it is easy to test
gravity, it would be more difficult to detect a new scalar force or a
new vector force such as those in \cite{KMY} and \cite{torsion}.
\begin{center}
\hglue -1truecm \epsfysize=200pt \epsfbox{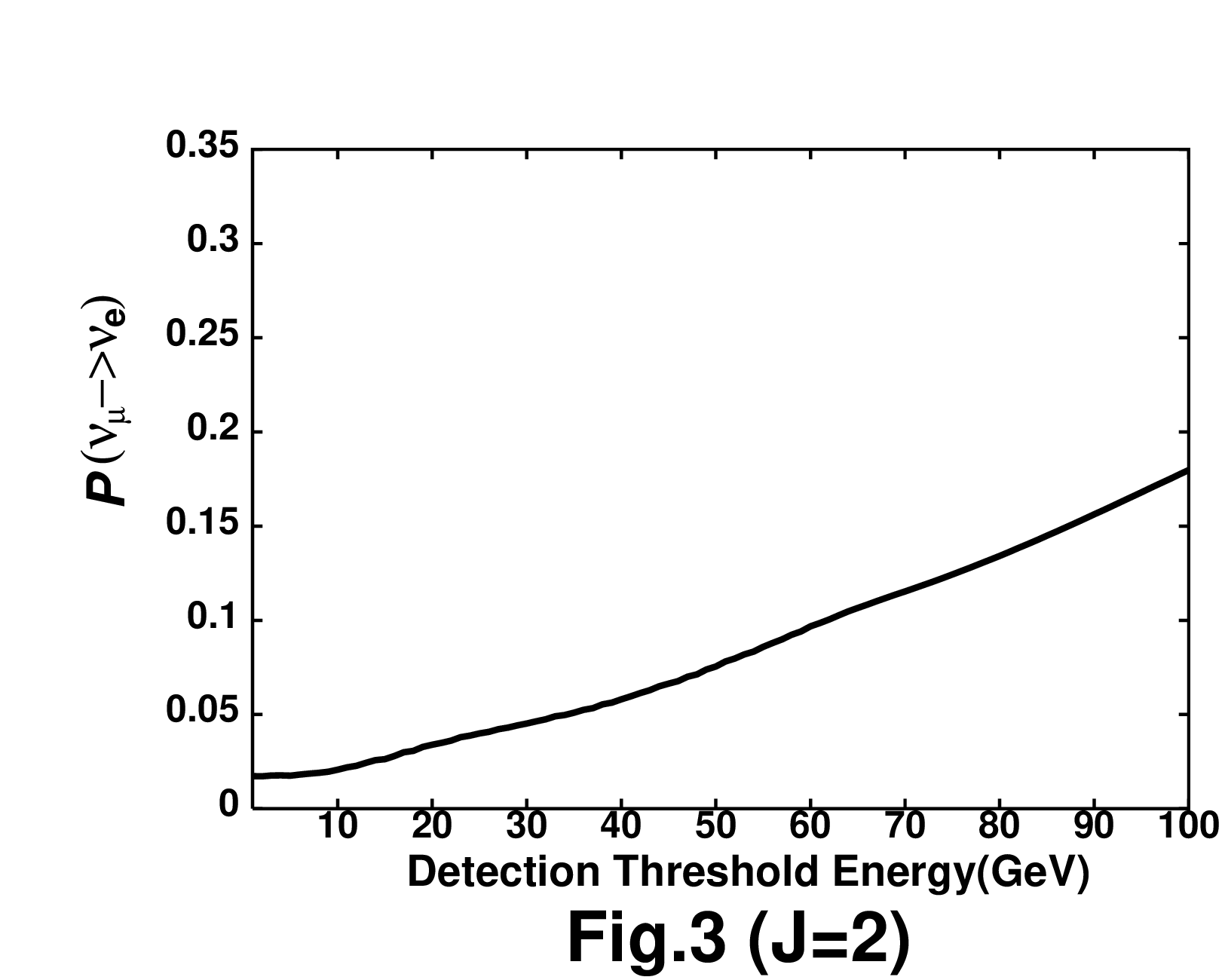}
\end{center}
\vglue 0.7truecm
\hglue -1truecm
\epsfysize=200pt \epsfbox{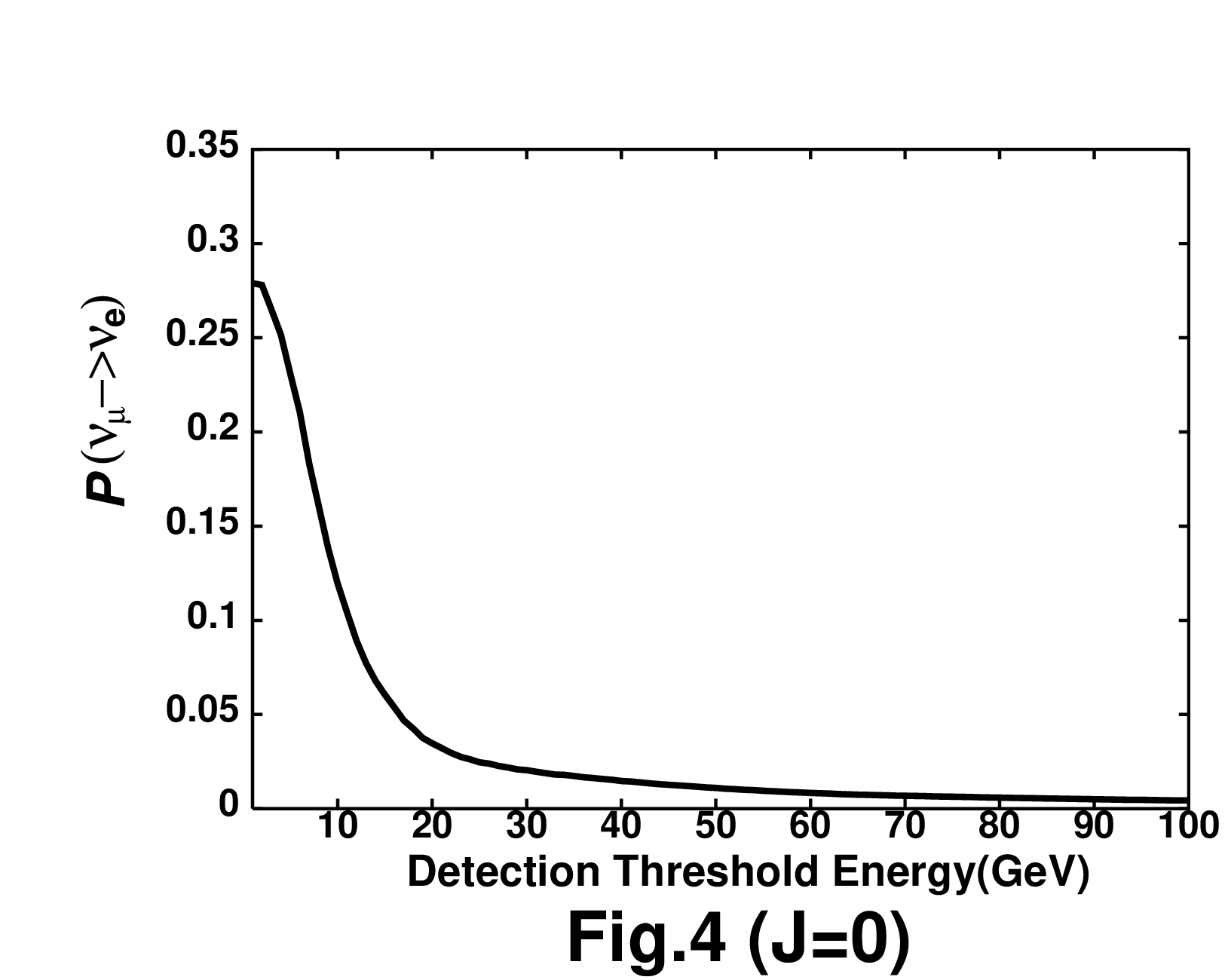}
\epsfysize=200pt \epsfbox{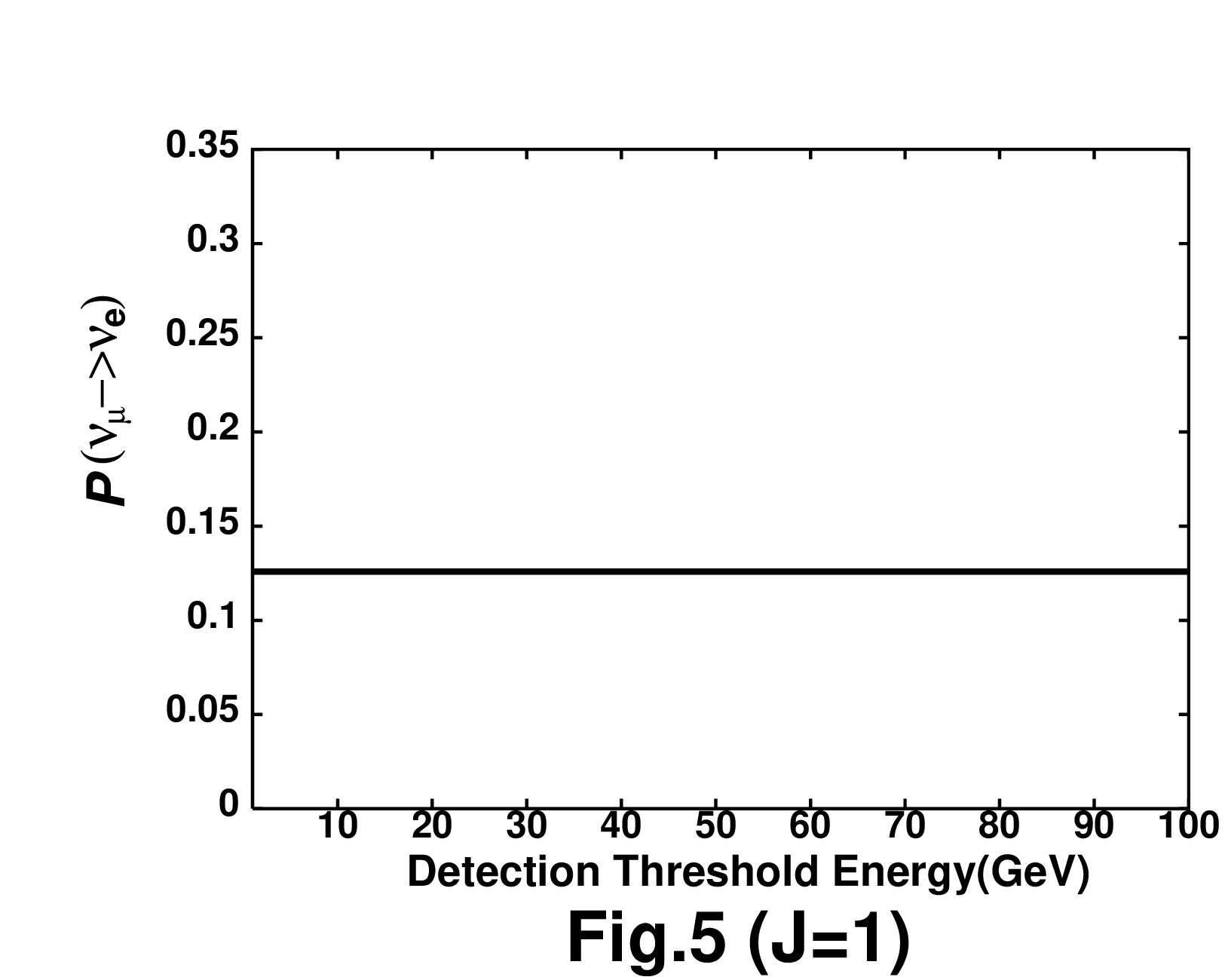}

\newpage

In this talk we have proposed long-baseline experiments to test the
universality of the gravitational couplings of neutrinos, and we found
that we could probe the dimensionless parameter $\Delta f$ as small as
$10^{-14}$ or $10^{-15}$ which is smaller by a few orders of
magnitudes than the upper limit on a breakdown of the equivalence
principle from different types of experiments.  Although we have not
evaluated systematic errors in detail, we hope our analysis will
stimulate and motivate long-baseline experiments in the near future.


\begin{thebibliography}{99}
\bibitem{Gasperini} M. Gasperini, Phys. Rev. {\bf D38} (1988) 2635; {\it ibid}
{\bf D39} (1989) 3606.
\bibitem{Halprin} A. Halprin and C.N. Leung, Phys. Rev. Lett. {\bf 67} (1991)
1833.
\bibitem{IMY} K. Iida, H. Minakata and O. Yasuda, Mod. Phys. Lett. {\bf A8}
(1993) 1037.
\bibitem{IMY2} K. Iida, H. Minakata and O. Yasuda, in preparation.
\bibitem{PHL} J. Pantaleone, A. Halprin and C.N. Leung, Phys. Rev. {\bf D47}
(1993) 4199.
\bibitem{Gravitation} C.W. Misner, K.S. Thorne and J.A. Wheeler,
``Gravitation'', W.H. Freeman and Company (1973).
\bibitem{Will} C.M. Will, Phys. Rep. {\bf 113} (1984) 345; Int. J. Mod. Phys.
{\bf D1} (1992) 13.
\bibitem{Longo} M.J. Longo, Phys. Rev. Lett. {\bf 60} (1987) 173; L.M. Krauss
and S. Tremaine, {\it ibid.} 176.
\bibitem{Stacey} F.D. Stacey, ``Physics of the Earth'', 2nd ed.,
John Wiley \& Sons, Inc. (1977).
\bibitem{Injector} R. Bernstein {\it et al.}, Conceptual Design Report:
Main Injector Neutrino Program, Fermilab, June 1991.
\bibitem{Moller} C. M\o ller, ``The Theory of Relativity'', 2nd ed.,
Oxford University Press (1972).
\bibitem{MSW} S.P. Mikheyev, A. Yu. Smirnov,
Sov. J. Nucl. Phys. {\bf 42} (1985) 913;
L. Wolfenstein,  Phys. Rev. {\bf D17} (1987) 2369
\bibitem{KMY} M. Kawasaki, H. Murayama and T. Yanagida, Mod. Phys. Lett. {\bf
A7} (1992) 563.
\bibitem{torsion} V. De Sabbata and M. Gasperini, Nuovo Cim. 65 {\bf A} (1981)
479.
\end{thebibliography}
\end{document}